\documentclass[11pt,a4paper,twocolumn,nofootinbib]{revtex4}

\usepackage[cp1250]{inputenc}
\usepackage{graphicx} 
\usepackage{subfigure}
\usepackage{geometry}
\usepackage{amsmath}
\usepackage{amssymb}
\usepackage{bm}
\usepackage{color}

\newgeometry{tmargin=2cm, bmargin=2cm, lmargin=2cm, rmargin=2cm}

\def\be{\begin{equation}}
\def\ee{\end{equation}}
\def\bea{\begin{eqnarray}}
\def\eea{\end{eqnarray}}

\def\II{\mathbb{I}}

\newcommand{\ie}{\textit{i.e. }}

\newcommand{\nocontentsline}[3]{}
\newcommand{\tocless}[2]{\bgroup\let\addcontentsline=\nocontentsline#1{#2}\egroup}

\renewcommand{\leq}{\;\leqslant\;}                   

\def\A{\bm{A}}

\begin{document}

\title{Detectability of Macroscopic Structures\\in Directed Asymmetric Stochastic Block Model}
\date{\today}
\author{Mateusz Wilinski\textsuperscript{1}}
\email{mateusz.wilinski@sns.it}
\author{Piero Mazzarisi\textsuperscript{2}}
\author{Daniele Tantari\textsuperscript{1}}
\author{Fabrizio Lillo\textsuperscript{2}}
\affiliation{\textsuperscript{1}Scuola Normale Superiore, Piazza dei Cavalieri 7, 56126 Pisa, Italy}
\affiliation{\textsuperscript{2}Dipartimento di Matematica, Porta di Piazza San Donato 5, 40126 Bologna, Italy}

\begin{abstract}
We study the problem of identifying macroscopic structures in networks,  characterizing the impact of introducing link directions on the detectability phase transition.
To this end, building on the stochastic block model, we construct a class of non trivially detectable directed networks.
We find closed form solutions by using belief propagation method showing how the transition line depends on the assortativity and the asymmetry of the network.
Finally, we numerically identify the existence of a hard phase for detection close to the transition point.
\end{abstract}

\maketitle

\section{INTRODUCTION}

Networks play a crucial role in the study of many complex systems, since
a large variety of such systems, derived from different scientific fields, are naturally described with nodes and edges.
A particularly interesting problem is the identification of network macroscopic structures, such as communities, multi-partite organization, core-periphery structures, etc. These are characterized by the fact that nodes can be divided into groups in such a way that nodes in the same group have the same linking attitude toward nodes of the same or of different groups.
The community organization is a special case observed when each node prefers to connect with nodes of its group \cite{fortunato2010community,danon2005comparing}.

Recently, there is a growing interest in understanding the limitations of community detection \cite{reichardt2008detectable,decelle2011inference,decelle2011asymptotic,nadakuditi2012graph} and interpreting results obtained for empirical networks \cite{peel2017ground}.
To this end, considering classes of networks, one looks for \emph{difficult} cases where the community structure is undetectable.
The notion of detectability threshold was first introduced for sparse graphs with block structure of connections in \cite{reichardt2008detectable} and the results were later found using Bayesian inference \cite{decelle2011inference,decelle2011asymptotic} and Random Matrix Theory \cite{nadakuditi2012graph}.
The work was also extended to dynamic case \cite{ghasemian2016detectability}, bipartite structures \cite{larremore2014efficiently} and weighted networks \cite{shi2018weighted}.
Detectability threshold refers to a sharp transition in parameters space separating two distinct regimes: (i) when it is possible, at least in part, to infer communities of nodes and (ii) when inference cannot work better than random assignments.
This intriguing phase transition is of practical importance and sheds a new light on the problem of community detection algorithms optimality.
A rigorous proof of the advocated phenomenon in static and symmetric case was given by 
\cite{mossel2015reconstruction,bordenave2015non}.
A recent summary on the subject can be found in \cite{abbe2018community}.
One should also remember that even though the results were shown for block models, the conclusions are more universal \cite{young2018universality}.

Simple graphs are often just an approximation of real-world networks.
Neglecting weights, multiple links or self loops may be justified in some cases but it also implies a loss of information.
An important generalization of a simple graph is a directed network.
Community detection in such networks is already an active field \cite{malliaros2013clustering}.
There is, however, no work, at least to our best knowledge, that describes the impact of introducing directions on the detectability phase transition.
In this paper we fill this gap by introducing belief propagation technique \cite{pearl1982reverend,mezard2001bethe} to the directed case of the stochastic block model \cite{holland1983stochastic,wang1987stochastic}.
As in the standard symmetric case, we first identify an entire class of non trivially detectable models, where both the average in- and out-degree are the same across all groups.
Later, we will concentrate on the case of many groups, where varying the assortativity leads to a hard detectable phase and a first order phase transition.
We will show how the asymmetry affects the size of this hard phase and the existence of discontinuity. Finally, as expected, we show that by introducing asymmetry in the affinity matrix, we are able to significantly increase the range of the detectable phase.

\section{MODEL}

\emph{Directed Stochastic Block Models (DiSBM)}.
We consider a network of $N$ nodes.
Links between nodes can be described by the $N \times N$ adjacency matrix $\A$ whose generic element $A_{ij}$ takes value $1$ if a link goes from $i$ to $j$, $0$ otherwise.
Since the network is directed, each couple of nodes may share two links, one for each direction, and, in general, the adjacency matrix is not symmetric.
 
A random directed graph is generated according to DiSBM as follows. Each node $i$ has a label $t_i\in\{1,2,...,q\}$ indicating which group it belongs to among $q$ possibilities.
Node labels are chosen independently, where for each node $i$ the probability that $t_i=a$ is $n_a$.
Once the groups are defined, for each pair of ordered nodes $(i,j)$, we put a link from $i$ to $j$ ($A_{ij}=1$) with probability $p_{t_it_j}$, where $p_{ab}$ are the entries of the $q\times q$ \emph{affinity matrix} $\boldsymbol{p}$. We are interested in the sparse regime where,  \ie $p_{ab}=O(1/N)$.
In this case, let us define the rescaled affinity matrix $c_{ab}=Np_{ab}$.
The total average degree of the graph is then $c=2\sum_{ab}c_{ab}n_an_b$, whereas the total average degree of a generic node in the group $a$ is $c_a=\sum_{b}(c_{ab}+c_{ba})n_b$, with $c^{out}_a=\sum_{b}c_{ab}n_b$ and $c^{in}_a=\sum_{b}c_{ba}n_b$ being respectively the average out- and in-degree in the same group.
Let $N_a$ be the number of nodes in the group $a$ and $M_{ab}$ the number of links from the group $a$ to the group $b$ for a specific graph realization.
Then $\lim_{N\to\infty}N_a/N=n_a$ and $\lim_{N\to\infty}NM_{ab}/N_a N_b=c_{ab}$.

\emph{Inference with Belief Propagation}.
In the detection setting, the adjacency matrix $\boldsymbol{A}$ is the only information available to us and the goal is to infer the true group assignments $\{t_i\}$ together with the parameters $\boldsymbol{\theta}=(q,\boldsymbol{n},\boldsymbol{c})$ of the model. 
The posterior probability that the model parameters take the value $\boldsymbol{\theta}$, conditioned to the observed adjacency $\boldsymbol{A}$ is
\be\label{eq:free}
P(\boldsymbol{\theta}|\boldsymbol{A})=\frac{P(\boldsymbol{\theta})}{P(\boldsymbol{A})} \sum_{\boldsymbol{t}}P(\boldsymbol{A},\boldsymbol{t}|\boldsymbol{\theta}).
\ee
The sum is over all possible assignments $\boldsymbol{t}$, where $t_i\in \{1,\ldots,q\}$ for each node $i$.
The prior $P(\boldsymbol{\theta})$ includes all the information about the parameters that does not depend on the graph and that we assume agnostically uniform.
$P(\boldsymbol{A},\boldsymbol{t}|\boldsymbol{\theta})$ is the \emph{likelihood} of a graph $\boldsymbol{A}$ with community assignments $\boldsymbol{t}$ according to DiSBM.
Analogously, assuming that we know, or have learned, the parameters $\boldsymbol{\theta}$, the posterior probability that the assignments take the value $\boldsymbol{t}$ conditioned to $\boldsymbol{A}$ and $\boldsymbol{\theta}$ is given by
\be
P(\boldsymbol{t} \vert \boldsymbol{A},\boldsymbol{\theta}) = \frac{P(\boldsymbol{A}, \boldsymbol{t} \vert \boldsymbol{\theta})}{\sum_{\boldsymbol{a}} P(\boldsymbol{A}, \boldsymbol{a} \vert \boldsymbol{\theta})}.
\ee
To maximize the number of correctly assigned labels we need to compute the marginals $\psi^i(a)=\sum_{\{t_j\}_{j\neq i}} P(\{t_j\}_{j\neq i},t_i=a \vert \boldsymbol{A},\boldsymbol{\theta})$ for each node $i$ of the posterior \cite{iba1999nishimori}, from which the most probable assignment is
\be \label{estimator}
\hat{t}_i=\operatorname{argmax}_{a\in\{1,\ldots,q\}} \psi^i(a).
\ee
Generalizing results from \cite{decelle2011asymptotic}, marginals can be estimated in a \emph{Belief Propagation} approach, using messages from neighbors as \begin{equation}\label{BPmarg}
\psi^i(a) = \frac{n_{a} e^{-h_{a}}}{Z^i} \prod_{k \in \partial i} \sum_{b} c_{ab}^{A_{ik}} c_{ba}^{A_{ki}} \psi^{k \rightarrow i}(b) ,
\end{equation}
where $h_{a} = \frac{1}{N} \sum_k \sum_{b} (c_{ba} + c_{ab}) \psi^k(b)$ and the messages $\psi^{i \rightarrow j}(a)$, for generic neighbors $(i,j)$ and $a\in\{1,\ldots,q\}$, are the fixed point of the  BP equations 
\be\label{BPmess}
\psi^{i \rightarrow j} (a) = \frac{n_{a} e^{-h_{a}}}{Z^{i \rightarrow j}} \prod_{k \in \partial i \setminus j} \sum_{b} c_{ab}^{A_{ik}} c_{ba}^{A_{ki}} \psi^{k \rightarrow i}(b)
\ee
The terms $Z^i$ and $Z^{i\to j}$ are normalization factors, since $\psi^i(a)$ and  $\psi^{i \rightarrow j} (a)$ are probabilities.
The neighborhood $\partial i$ of a node $i$ is intended as those nodes for which $A_{ij}=1$ or $A_{ji}=1$.
Thus it is important to stress that message information propagates along the undirected skeleton of the network, while directions only change the type of information transmitted. 
Marginals and messages are used also to maximize $(\ref{eq:free})$, finding the optimal parameters
\begin{eqnarray}
n_a &=& \frac{\sum_i \psi^i(a)}{N},\label{nish1}\\
c_{ab} &=& \frac{1}{N} \frac{c_{ab}}{n_a n_b} \sum_{(i, j) \in E}  \frac{\psi^{i \rightarrow j}(a) \cdot \psi^{j \rightarrow i}(b)}{Z^{ij}},\label{nish2}
\end{eqnarray}
where $E$ is the set of all, directed edges.
We use an Expectation-Maximization procedure of alternatively solving until convergence the BP equations $(\ref{BPmarg})$-$(\ref{BPmess})$ and the equations $(\ref{nish1})$-$(\ref{nish2})$.
More details regarding the messages form and inferring the parameters, can be found in the appendix.

To measure the accuracy of the estimator $(\ref{estimator})$ we introduce the \emph{overlap} between planted and inferred assignments as
\be
Q = \frac{\max_{\pi} \frac{1}{N} \sum_{i=1}^N \delta(\pi(\hat{t}_i), t_i) - \max_a n_a}{1 - \max_a n_a},
\ee
where  $\pi$ iterates the permutations on $q$ elements and $\delta(\cdot,\cdot)$ is the Kronecker delta.
$Q=1$ for perfect matching while $Q=0$ when all nodes are assigned to the largest group (for equally sized groups $Q=0$ when community inference performs as a random guess).

\begin{figure}
\begin{center}
\includegraphics[width=0.5\textwidth]{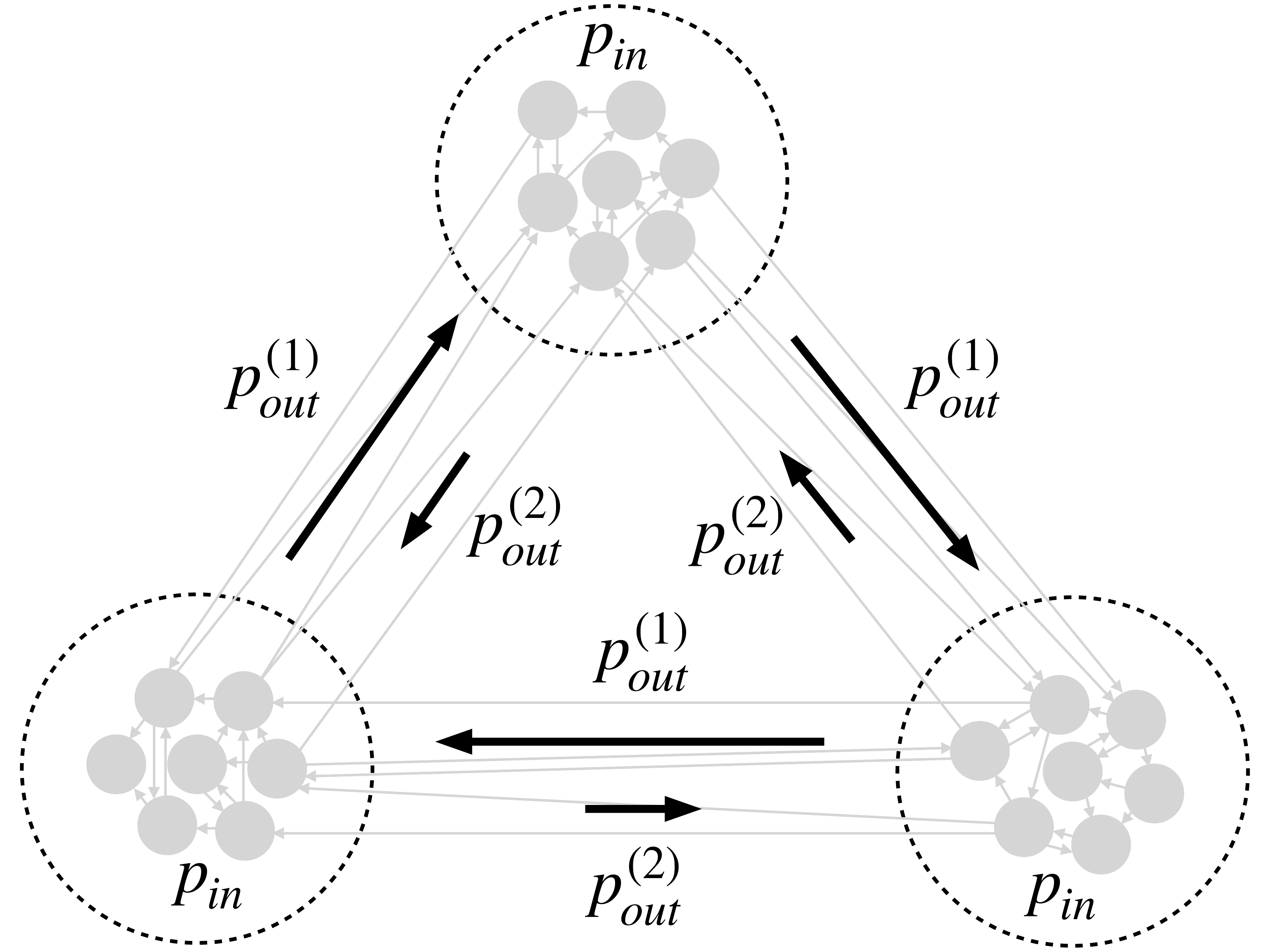}
\caption{Schematic plot showing a network generated with asymmetric planted partition in case of $q=3$.
Probabilities of given edges are described by $p_{\text{in}}$, $p^{(1)}_{\text{out}}$ and $p^{(2)}_{\text{out}}$.}
\label{fig:example}
\end{center}
\end{figure}

As highlighted in \cite{decelle2011asymptotic} for the case of undirected graphs, the problem of community detection is trivial when different groups have different total average degree $c_a$. In this case the total degree distribution is multimodal (linear combination of Poisson distributions with different means) and nodes can be classified according to their degree, since each mode of the distribution is correlated with the corresponding group.
The addition of directions make the detection even easier, since nodes can be classified according to their in(out)-degree even when the groups have homogeneous total average degree.
For example an affinity matrix $\begin{pmatrix}c/2 & c\\0 & c/2\end{pmatrix}$ generates a directed network with a strong bipartite structure that becomes completely undetectable without the directions.
Thus the problem becomes non trivial once
\be \label{eq:cond}
c_a^{in}=c^{out}_a=c_a/2=c/2, \ \ \ \forall a=1,\ldots, q.
\ee
\emph{Circulant models}.
Without losing generality we consider the case of equally sized group, i.e. $n_a=1/q$, for which the conditions $(\ref{eq:cond})$ means that the affinity matrix must be a multiple of a \emph{doubly stochastic} matrix, i.e. with constant sums along rows and columns.
This implies it can be represented as a linear combination of permutation matrices, i.e. there exist coefficients $c_{in},c^{(1)}_{out},\ldots, c^{(k)}_{out}$ and permutations $\pi_1,\ldots,\pi_k$, $\pi_i:\{1,\ldots,q\}\to \{1,\ldots,q\}$ such that
\be
\boldsymbol{c}=c_{in}\II+c^{(1)}_{out}P^{\pi_1}+\ldots+ c^{(k)}_{out}P^{\pi_k},
\ee
where $P^\pi_{ij}$ is defined as $1$ if $j=\pi(i)$ and $0$ otherwise.
Of particular interest is the case when the permutations $\pi$ are cycles of different order $\mathcal{C}_k(i)=(i+k$ mod $q) + 1$ because they bring to asymmetric affinity matrices where the role of the directions emerge.
They produce the class of non trivial directed models where the affinity matrix $\boldsymbol{c}=c_{in}\II+c^{(1)}_{out}P^{\mathcal{C}_1}+\ldots+ c^{(k)}_{out}P^{\mathcal{C}_k}$ is a circulant matrix, i.e. with constant diagonals of any order.
A special, yet important, case is what we call \emph{Asymmetric Planted Partition} model, in which the affinity matrix depends on just three parameters $c_{in},c^{(1)}_{out}, c^{(2)}_{out}$ as
\be
\bm{c} = \begin{pmatrix} 
    c_{\text{in}} & c^{(1)}_{\text{out}} & c^{(2)}_{\text{out}} & \dots & c^{(2)}_{\text{out}} \\
    c^{(2)}_{\text{out}} & c_{\text{in}} & c^{(1)}_{\text{out}} & \dots & c^{(1)}_{\text{out}} \\
    c^{(1)}_{\text{out}} & c^{(2)}_{\text{out}} & c_{\text{in}} & \dots & c^{(2)}_{\text{out}} \\
    \vdots & \vdots &  \vdots & \ddots & \vdots \\
    c^{(1)}_{\text{out}} & c^{(2)}_{\text{out}} & c^{(1)}_{\text{out}} & \dots & c_{\text{in}}
    \end{pmatrix}.
\label{eq:part}
\ee
A schematic representation of the simplest case of a $q=3$ network generated by such a model is shown in Fig. \ref{fig:example}.
Note that from the perspective of asymmetry in the model, it is only interesting to analyze the above case when $q$ is odd.
Otherwise, we end up with a symmetric affinity matrix and the detectability is described only by the network assortativity.
Let us define
\be
\varepsilon = \frac{c^{(1)}_{\text{out}} + c^{(2)}_{\text{out}}}{2 c_{\text{in}}}, \quad \quad \gamma = \frac{c^{(1)}_{\text{out}}}{c^{(2)}_{\text{out}}},
\ee
to measure, respectively, the level of assortativity, similarly to the undirected case \cite{decelle2011inference,decelle2011asymptotic}, and the level of asymmetry in the direction of connections between each pair of groups.

\section{RESULTS}

\emph{Detectability transitions}.
When Eq. (\ref{eq:cond}) holds, the BP equations always have the so called paramagnetic fixed point $\psi^{i\to j}_{t_i}=n_{t_i}$, which does not carry any information on the group structure.
When the paramagnetic solution is locally stable the inference becomes impossible or at best extremely hard, as solving the hardest known optimization problems \cite{krzakala2009hiding,franz2001ferromagnet}.
To check the stability we study how a small random perturbations of the paramagnetic fixed point propagates as the BP equations are iterated.
To do this we use the tree like approximation, which is justified in the sparse network regime.
Let us consider the path from a leaf $k_d$ to a root $k_0$ to be $k_d, k_{d-1}, \dots, k_1, k_0$ and assume that on the leaves the paramagnetic fixed point is perturbed as 
\begin{equation}
\psi^{k_d}_{t}=n_{t}+\epsilon^{k_d}_t.
\end{equation}
Following the idea of leaves perturbation, taken from the undirected case \cite{decelle2011asymptotic}, we calculate the transfer matrix
\begin{equation}
\begin{split}
T_i^{ab} &= \left. \frac{\partial \psi_a^{k_i}}{\partial \psi_b^{{k_{i+1}}}} \right|_{\psi_t = n_t} \\
&= n_a \left( \frac{c_{ab}^{A_{k_i,k_{i+1}}} \cdot c_{ba}^{A_{k_{i+1},k_i}}}{c / 2} - 1 \right).
\end{split}
\end{equation}
Unlike in the undirected case, the transfer matrix does depend on $i$.
Nevertheless, if we assume that two way links are neglectable, we only obtain either $T_{ab} = n_a \left( \frac{c_{ab}}{c / 2} - 1 \right)$ or its transposition $T^T_{ba}= n_a \left( \frac{c_{ba}}{c / 2} - 1 \right)$, depending on the direction of the link.
The perturbation $\epsilon_{t_0}^{k_0}$ on the root due to the perturbation $\epsilon_{t_d}^{k_d}$ on the leaf $k_d$ can then be written in matrix notation as
\begin{equation}
\epsilon^{k_0} := T_{k_d\to k_0}  \epsilon^{k_d}= \left( \prod_{i=0}^{d-1} T_i \right) \epsilon^{k_d}.
\end{equation}
If we consider the total perturbation induced to the root by all the leaves at distance $d$ on the tree we obtain
\begin{equation}
\epsilon^{k_0} := \sum_{k_d} T_{k_d\to k_0}  \epsilon^{k_d},
\end{equation}
whose strength, if we consider uncorrelated perturbations on the leaves, reads as
\begin{equation}
\begin{split}
\langle\epsilon^{k_0},\epsilon^{k_0}\rangle &= \sum_{k_d} \langle \epsilon^{k_d},T_{k_d\to k_0}  ^T T_{k_d\to k_0}  \epsilon^{k_d}\rangle \\
&\leq \sum_{k_d} \lambda_{k_d} \langle\epsilon^{k_d},\epsilon^{k_d}\rangle,
\end{split}
\end{equation}
where $\lambda_{k_d}$ is the largest eigenvalue of the matrix $T_{k_d\to k_0}  ^T T_{k_d\to k_0}$.
In principle $\lambda_{k_d}$ does depend on the path, \textit{i.e.} the leaf $k_d$.
However, as soon as $[T,T^T]=0$ we have that $T_{k_d\to k_0}^T T_{k_d\to k_0}= (T^T T)^d$, independently from $k_d$.
As a result, assuming normalized perturbations on the leaves, we get
\begin{equation}
\langle\epsilon^{k_0},\epsilon^{k_0}\rangle \leq \lambda^d \sum_{k_d} \langle\epsilon^{k_d},\epsilon^{k_d}\rangle \sim \lambda^d c^d,
\end{equation}
where $\lambda$ is the largest eigenvalue of $T^TT$ and  $c^d$ is the expected number of leaves at distance $d$. 
This leads to a following detectability threshold:
\be\label{eq:lambda_cond}
c \lambda  = 1,
\ee
As soon as $c \lambda < 1$, the perturbation vanishes as we go through the network and the paramagnetic fixed point is locally stable: it can either be globally attractive (paramagnetic/undetectable phase, where the network is indistinguishable from an Erdos-Renyi graph) or  coexist with solutions correlated with the original assignments (hard phase).
On the other hand, for $c \lambda  > 1$ the perturbation is amplified exponentially: the paramagnetic fixed point is unstable and BP easily infers the  original communities (ordered/detectable phase).
In the special case described with Eq. (\ref{eq:part}) the detectability condition is as follows:
\be\label{eq:max_lambda}
c q^2 < (2 c_{\text{in}} - (c^{(1)}_{\text{out}} + c^{(2)}_{\text{out}}))^2 + (c^{(1)}_{\text{out}} - c^{(2)}_{\text{out}})^2 \cot^2\left( \frac{\pi}{2q} \right).
\ee
The first term in the r.h.s. represents the signal carried by the symmetrized adjacency matrix, thus 
for $c^{(1)}_{\text{out}} = c^{(2)}_{\text{out}}$ we get the original equation for the largest eigenvalue in the undirected case of planted partition model \cite{decelle2011asymptotic}. Conversely, the second term represents the signal carried by the directions because of the asymmetry in the affinity matrix.

\begin{figure*}
\begin{center}
\includegraphics{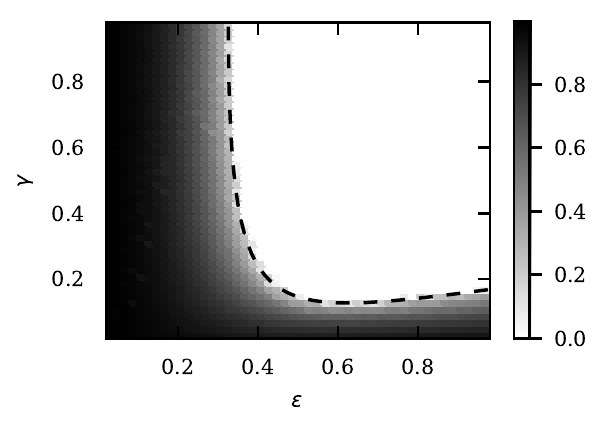}
\includegraphics{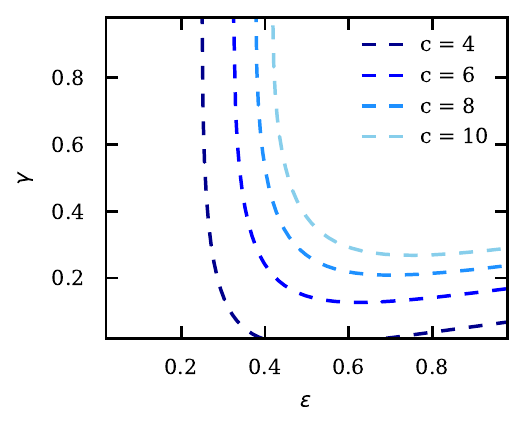}
\includegraphics{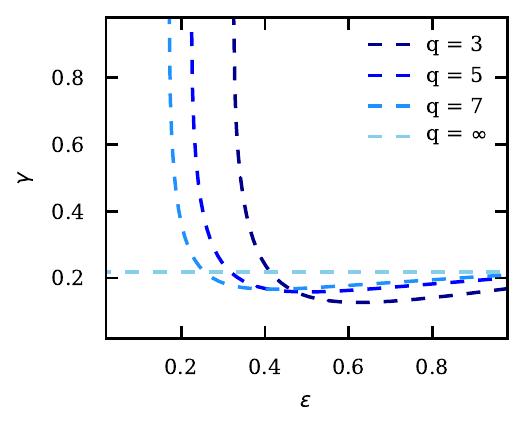}
\caption{Left: Two dimensional phase diagram of an asymmetric planted partition model with $q = 3$ groups and average degree $c = 6$, obtained numerically \cite{code} for network with $N = 9 \cdot 10^4$ nodes.
The colour scale corresponds to the overlap between the inferred and the real nodes' assignment.
Dashed black line represents the analytical approximation of Eq. (\ref{eq:max_lambda}).
Middle: analytical critical lines when number of groups equal to $q = 3$ and varying average degree $c$.
Right: analytical critical lines when network average degree is equal to $c = 6$ and varying number of groups $q$.}
\label{fig:2d}
\end{center}
\end{figure*}

For low number of groups,  for example $q = 3$ in the left panel of Fig. \ref{fig:2d}, there are only two distinct phases, the ordered phase with positive overlap and the paramagnetic phase with zero overlap, and the observed phase transition is continuous at the critical line described by Eq. $(\ref{eq:max_lambda})$.
In this case BP recursion converges quickly to the globally attractive fixed point (many equivalent up to a permutation in the ordered phase) for any random initial conditions for messages and marginals: convergence time diverges at the transition.
Interestingly, one can reach the detectable phase by either increasing the assortativity $\varepsilon$ or by decreasing the symmetry $\gamma$.
In both cases there is a range of values that guarantees detectability, regardless of the other parameter value.
It should also be pointed out that the case of $\gamma = 1$ is equivalent to the undirected case.
As a result, one can compare different values of $\gamma$ and see the actual result of taking into account asymmetry of connections.

From Eq. (\ref{eq:max_lambda}) one gets that increasing the average degree decreases the size of the paramagnetic (i.e. undetectable) phase.
As shown on the middle panel of Fig. \ref{fig:2d}, this behaviour is the same to all effects from assortativity and connections symmetry perspective.
The right panel shows a different behaviour when varying the number of groups.
The critical value of $\varepsilon$, below which the network is detectable regardless of $\gamma$, decreases by increasing $q$.
This is somehow intuitive because higher number of groups increases the task complexity.
Surprisingly, however, the critical value of $\gamma$, below which the network is detectable regardless of $\varepsilon$, does not change much.
Moreover, although the change is small, this value is increasing, making even more important not to disregard the information about edge direction.
When $q \rightarrow \infty$ the assortativity becomes irrelevant and one finds that there is only critical value $\gamma_c = (\frac{2}{\pi} \sqrt{c} - 1) / (\frac{2}{\pi} \sqrt{c} + 1)$.

\begin{figure*}
\begin{center}
\includegraphics{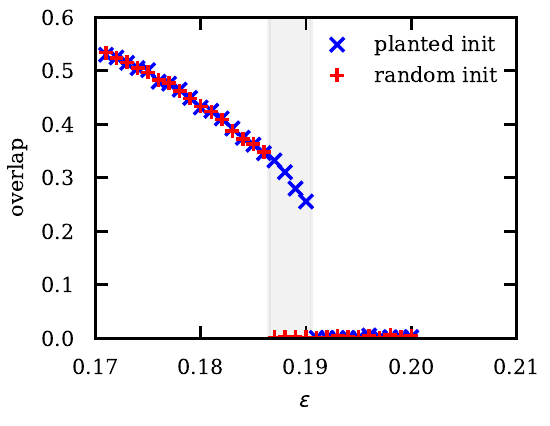}
\includegraphics{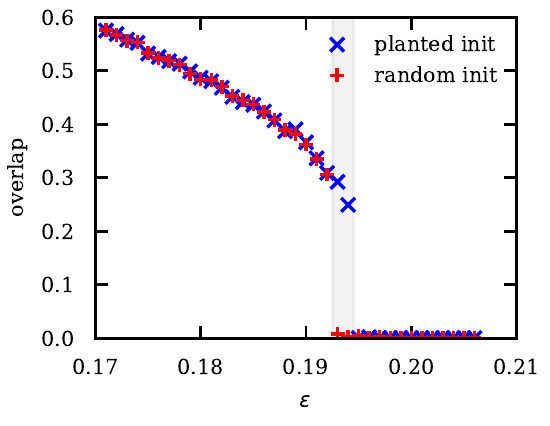}
\includegraphics{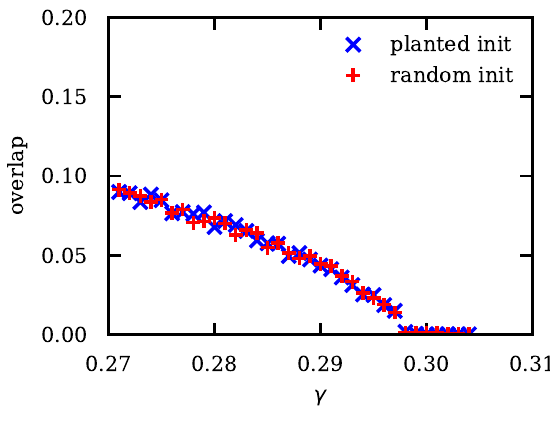}
\caption{Phase diagrams of an asymmetric planted partition with $q = 9$ groups and average degree $c = 9$.
Results obtained numerically \cite{code}, for networks with $N = 3.6 \cdot 10^5$ nodes, using belief propagation with random initial conditions (red pluses) and initiated with the correct assignments (blue crosses). 
The great area approximates the hard phase regime, according to numerical results.
Left: Results for fixed asymmetry parameter $\gamma = 0.8$.
Middle: Results for fixed asymmetry parameter $\gamma = 0.6$.
Right: Results for fixed assortativity parameter $\epsilon = 0.8$.}
\label{fig:dis}
\end{center}
\end{figure*}

Similarly to the undirected case \cite{decelle2011inference,mezard2006reconstruction}, increasing the number of groups above $q = 4$ leads to the appearance of the hard phase.
Moreover, it changes the order of the transition from second to first, which means that there is a discontinuous jump in the overlap.
The hard phase can be identified by studying the sensitivity of BP fixed points to initial conditions: in the hard phase the fixed point correlated with the original assignment coexists with the paramagnetic one and has a very small basin of attraction.
Thus, starting the recursion from random messages and marginals, BP will never reach a solution with positive overlap.
Nevertheless such a solution does exist and still can be found by initialising marginals and messages close enough to it, for example $\psi^i(a)=\psi^{i\to j}(a)=\delta_{a,t_i}$.
The left and the middle panels of Fig. \ref{fig:dis} indicate that, in this case, not only the transition point moves towards lower assortativity with increasing asymmetry, but also the size of the hard phase is decreasing.
Furthermore, looking at the phase diagram for the asymmetry parameter, (right panel of Fig. \ref{fig:dis}) we find a different behaviour.
Although for $\epsilon = 0.8$ we can still observe a small jump in the overlap, the presence of the hard phase cannot be confirmed with numerical simulations.
When the assortativity disappears completely, the jump seemingly vanishes as well.

\section{Conclusions}

This paper studies the limits of community detection in the directed stochastic block model with an asymmetric affinity matrix.
We showed the detectability condition for a broad range of cases with commuting transfer matrices.
We proposed and focused on the asymmetric planted partition and obtained the detectability threshold as a function of its parameters.
The results show that correlation with the correct assignments can be achieved not only by increasing assortativity but also by increasing asymmetry.
Both parameters, above a given threshold, lead to positive detectability regardless of the value of the other.
This was also confirmed by extensive numerical simulations.
Moreover, similarly to the undirected case, the phase transition type depends on the number of groups.
For small number of groups we observe a first order phase transitions.
When the number of groups is exceeds four, the phase transitions becomes discontinuous.
The observed jump is, however, decreasing with increasing asymmetry and so is the size of the hard phase.
Furthermore, for high asymmetry it becomes indistinguishable from a continuous phase transition.
Varying the number of groups also affects the critical line.
Larger number of groups require a higher assortativity level in order for the network to have detectable community structure.
Interestingly, it does not affect the critical level of symmetry much.
A more intuitive effect is observed when we vary the average degree.
In this case, both the critical assortativity and dissymmetry do increase with increasing degree.

Our, purely theoretical, work has important implications on the study of empirical networks.
Majority of real-world networks are directed but researchers tend to neglect that and use the undirected projection.
As a result, they loose an important information which, as shown in this article, may affect the detectability properties of a network.
In the worst case scenario, which is not very likely but still possible, the network may even shift from the detectable phase toward the undetectable regime.
Otherwise, using undirected projection will result in decreasing the achievable correlation with the actual assignments, when trying to infer them.
Importantly, the process is irreversible without getting back to the directed representation of a given network.
In future work, we plan to analyse empirical networks and show how big is the effect of neglecting directions.
Broader question that emerges from the analysed problem is to what extend information that are often neglected in network analysis, like directions, weights etc., do affect community detection or other, similar tasks.

\begin{acknowledgments}
DT acknowledges GNFM-Indam and Scuola Normale Superiore for financial support from the project SNS18\_A\_TANTARI.
\end{acknowledgments}

\bibliography{bibliography}

\appendix

\section{Parameters and Assignments Inference}

The probability that a directed stochastic block model, described with parameters $\boldsymbol{\theta} = \{ q, \{n_a\}, \{ p_{ab} \} \}$, generates a graph $G$, with adjacency matrix $\boldsymbol{A}$, is equal to
\begin{equation}
P(\boldsymbol{A}, \boldsymbol{t} \vert \boldsymbol{\theta}) = \prod_{i \neq j} \left( p_{t_i, t_j}^{A_{ij}} (1 - p_{t_i, t_j})^{1-A_{ij}} \right) \prod_i n_{t_i}.
\end{equation}
Using Bayes' rule we can find a probability of a given assignment $\boldsymbol{t}$, as a function of the matrix $\boldsymbol{A}$ and model parameters $\boldsymbol{\theta}$
\begin{equation}
P(\boldsymbol{t} \vert \boldsymbol{A}, \boldsymbol{\theta}) = \frac{P(\boldsymbol{A}, \boldsymbol{t} \vert \boldsymbol{\theta})}{\sum_{\boldsymbol{r}} P(\boldsymbol{A}, \boldsymbol{r} \vert \boldsymbol{\theta})}.
\end{equation}
In the directed case above equations lead to belief propagation messages that differ from their undirected counterparts.
They need to take into account both incoming and outgoing links which results with following equations for messages and marginals accordingly:
\begin{widetext}
\begin{equation}
\psi_{t_i}^{i \rightarrow j} = \frac{n_{t_i}}{Z^{i \rightarrow j}} \prod_{\substack{k \neq i \\ k \neq j}} \left( \sum_{t_k} c_{t_i t_k}^{A_{ik}} c_{t_k t_i}^{A_{ki}} \left( 1 - \frac{c_{t_i t_k}}{N} \right)^{1 - A_{ik}} \left(1 - \frac{c_{t_k t_i}}{N} \right)^{1 - A_{ki}} \psi_{t_k}^{k \rightarrow i} \right),
\end{equation}
\begin{equation}
\psi_{t_i}^i = \frac{n_{t_i}}{Z^i} \prod_{k \neq i} \left( \sum_{t_k} c_{t_i t_k}^{A_{ik}} c_{t_k t_i}^{A_{ki}} \left( 1 - \frac{c_{t_i t_k}}{N} \right)^{1 - A_{ik}} \left( 1 - \frac{c_{t_k t_i}}{N} \right)^{1 - A_{ki}} \psi_{t_k}^{k \rightarrow i} \right),
\end{equation}
\end{widetext}
where $c_{ab} = N p_{ab}$.
For large sparse networks, when $N \gg 1$ and $c_{ab} = O(1)$ we can rewrite the message passing by neglecting terms of sub-leading order in $N$.
\begin{equation}
\psi_{t_i}^{i \rightarrow j} = \frac{n_{t_i} e^{-h_{t_i}}}{Z^{i \rightarrow j}} \prod_{k \in \partial i \setminus j} \left( \sum_{t_k} c_{t_i t_k}^{A_{ik}} c_{t_k t_i}^{A_{ki}} \psi_{t_k}^{k \rightarrow i} \right),
\end{equation}
where
\begin{equation}
h_{t_i} = \frac{1}{N} \sum_k \sum_{t_k} (c_{t_k t_i} + c_{t_i t_k}) \psi_{t_k}^k.
\end{equation}
Similarly for marginals
\begin{equation}
\psi_{t_i}^i = \frac{n_{t_i} e^{-h_{t_i}}}{Z^i} \prod_{k \in \partial i} \left( \sum_{t_k} c_{t_i t_k}^{A_{ik}} c_{t_k t_i}^{A_{ki}} \psi_{t_k}^{k \rightarrow i} \right).
\end{equation}
In this case the partition function can be written as a sum
\begin{equation}
Z^{i} = \sum_{t_i} n_{t_i} e^{-h_{t_i}} \prod_{k \in \partial i} \left( \sum_{t_k} c_{t_i t_k}^{A_{ik}} c_{t_k t_i}^{A_{ki}} \psi_{t_k}^{k \rightarrow i} \right).
\end{equation}
Moreover, we can write the free energy estimate for BP
\begin{equation}
\begin{split}
f_{\mathrm{BP}}(q, \{ n_a \}, \{ c_{ab} \}) &= \frac{1}{N} \sum_{(i,j) \in E} \log Z^{ij}\\
&- \frac{1}{N} \sum_i \log Z^i - \frac{c}{2},
\end{split}
\end{equation}
where $E$ is the set of all directed edges, which may include separately edges $(i,j)$ and $(j,i)$.
In the first term we have
\begin{equation}
Z^{ij} = \sum_{a,b} c_{ab} \, \psi_a^{i \rightarrow j} \, \psi_b^{j \rightarrow i} = \Psi^{i \rightarrow j} \, c \, \Psi^{j \rightarrow i}.
\end{equation}
Using this form of the free energy we can calculate the Nishimori conditions and obtain
\begin{equation}
n_a = \frac{\sum_i \psi_{a}^i}{N},
\end{equation}
\begin{equation}
c_{ab} = \frac{1}{N} \frac{c_{ab}}{n_a n_b} \sum_{(i,j) \in E} \frac{\psi_a^{i \rightarrow j} \cdot \psi_b^{j \rightarrow i}}{Z^{ij}}.
\end{equation}
This way, we can infer both the group assignments and the underlying directed stochastic block model parameters.

\section{Asymmetric Planted Partition Case}

Let us consider the asymmetric planted partition matrix, defined by a following affinity matrix:
\begin{equation}
\bm{p} = \begin{bmatrix} 
    p_{\text{in}} & p^{(1)}_{\text{out}} & p^{(2)}_{\text{out}} & \dots & p^{(2)}_{\text{out}} \\
    p^{(2)}_{\text{out}} & p_{\text{in}} & p^{(1)}_{\text{out}} & \dots & p^{(1)}_{\text{out}} \\
    p^{(1)}_{\text{out}} & p^{(2)}_{\text{out}} & p_{\text{in}} & \dots & p^{(2)}_{\text{out}} \\
    \vdots & \vdots &  \vdots & \ddots & \vdots \\
    p^{(1)}_{\text{out}} & p^{(2)}_{\text{out}} & p^{(1)}_{\text{out}} & \dots & p_{\text{in}}
    \end{bmatrix}
\end{equation}
and equal group fractions.
It is a special case of a circulant matrix, which is a matrix of a following form
\begin{equation}
S = \begin{bmatrix} 
    s_0 & s_1 & s_2 & \dots & s_{n-1} \\
    s_{n-1} & s_0 & s_1 & \dots & s_{n-2} \\
    s_{n-2} & s_{n-1} & s_0 & \dots & s_{n-3} \\
    \vdots & \vdots & \vdots & \ddots & \vdots \\
    s_1 & s_2 & s_3 & \dots & s_0 
    \end{bmatrix}
\end{equation}
As a result, both $T$ and $T^T T$ matrices are also circulant.
The transfer matrix, in this case, is parametrised by three values
\begin{equation}
s_k(T) = \begin{cases}
X & \text{if} \quad k = 0\\
Y & \text{if} \quad k \text{ odd}\\
Z & \text{if} \quad k \text{ even}
\end{cases}
\end{equation}
where $k = 0,1,2,\dots,q-1$ and
\begin{equation}
\begin{split}
X = \frac{1}{q} \left( \frac{c_{\text{in}}}{c/2} - 1 \right),\\
Y = \frac{1}{q} \left( \frac{c^{(1)}_{\text{out}}}{c/2} - 1 \right),\\
Z = \frac{1}{q} \left( \frac{c^{(2)}_{\text{out}}}{c/2} - 1 \right).
\end{split}
\end{equation}
We can use the above parameters to describe the matrix $T^T T$ as a circulant matrix
\begin{widetext}
\begin{equation}
s_k(T^T T) = \begin{cases}
X^2 + \frac{q-1}{2} (Y^2 + Z^2) & \text{if} \quad k = 0\\
\frac{k-1}{2} (Y^2 + Z^2) + (q-k-1) Y Z + X (Y + Z) & \text{if} \quad k \text{ odd}\\
\frac{q-k-1}{2} (Y^2 + Z^2) + (k-1) Y Z + X (Y + Z) & \text{if} \quad k \text{ even}
\end{cases}
\label{eq:TT}
\end{equation}
\end{widetext}
The next step is to find the largest eigenvalue of the above matrix.
For circulant matrices it is known that the eigenvalues are given by
\begin{equation}
\lambda_{m} = \sum_{k=0}^{q-1} s_k e^{-\frac{2 \pi i m k}{q}},
\end{equation}
where $m = 0,1,2,\dots,q-1$.
Since $s_k(T^T T) = s_{q-k}(T^T T)$, all the eigenvalues are real and we can concentrate on the real part of the above equation.
\begin{equation}
\lambda_{m} = \sum_{k=0}^{q-1} s_k \cos\left(2 \pi m \frac{k}{q}\right).
\label{eq:cir_eig}
\end{equation}
This form of the equation is still difficult though.
In order to sum the elements, we will use a simple trigonometric identity
\begin{widetext}
\begin{equation}
\cos\left( 2 \pi m \frac{k}{q} \right) \cos\left( 2 \pi m \frac{1}{2 q} \right) = \frac{1}{2} \cos\left( 2 \pi m \frac{2 k - 1}{2 q} \right) + \frac{1}{2} \cos\left( 2 \pi m \frac{2 k + 1}{2 q} \right).
\label{eq:id1}
\end{equation}
\end{widetext}
Combining Eq. (\ref{eq:TT}), (\ref{eq:cir_eig}) and (\ref{eq:id1}) leads to
\begin{widetext}
\begin{equation}
\begin{split}
\lambda_{m} &= X^2 + \frac{q-1}{2} (Y^2 + Z^2) + (q - 2) Y Z + X (Y + Z)\\
&+  \sum_{k=1}^{q-2} \left[ \frac{q - 2 \pm 1}{2} (Y^2 + Z^2) + (q - 2 \mp 1) Y Z + 2 X (Y + Z) \right] \cdot \frac{\cos\left(2 \pi m \frac{2 k + 1}{2 q}\right)}{2 \cos\left( \pi \frac{m}{q} \right)},
\end{split}
\label{eq:sim_eig1}
\end{equation}
\end{widetext}
where $\pm$ means $+$ when $k$ is odd and $-$ when $k$ is even, vice versa for $\mp$.
To eventually get rid off the dependence on the parity of $k$ we will need to use one more trigonometric identity
\begin{widetext}
\begin{equation}
\cos\left( 2 \pi m \frac{2 k + 1}{2 q} \right) \cos\left( 2 \pi m \frac{1}{2 q} \right) = \frac{1}{2} \cos\left( 2 \pi m \frac{k}{q} \right) + \frac{1}{2} \cos\left( 2 \pi m \frac{k + 1}{q} \right).
\label{eq:id2}
\end{equation}
\end{widetext}
Applying it to Eq. (\ref{eq:sim_eig1}) give
\begin{widetext}
\begin{equation}
\begin{split}
\lambda_{m} &= X^2 + \frac{q - 1}{2} (Y^2 + Z^2) + (q - 2) Y Z + X (Y + Z)\\
&+ \left(\frac{q - 3}{4} (Y^2 + Z^2) + \frac{q - 1}{2} \, Y Z + X (Y + Z)\right) \frac{\cos\left(2 \pi \frac{m}{q}\right)}{\cos^2\left( \pi \frac{m}{q} \right)}\\
&+  \left[ (q - 2) (Y^2 + Z^2) + 2 (q - 2) Y Z + 4 X (Y + Z) \right] \sum_{k=1}^{q-3} \frac{\cos\left(2 \pi m \frac{k + 1}{q}\right)}{4 \cos^2\left( \pi \frac{m}{q} \right)}.
\end{split}
\label{eq:sim_eig2}
\end{equation}
\end{widetext}
The final sum can be calculated using
\begin{equation}
\sum_{k=0}^{n-1} \cos( 2 \pi k) = 1,
\label{eq:id3}
\end{equation}
where $n$ is odd.
Finally, after simplifying all the parts we get
\begin{equation}
\begin{split}
\lambda_{m} = \frac{1}{c^2 q^2} &\left( (2 c_{\text{in}} - (c^{(1)}_{\text{out}} + c^{(2)}_{\text{out}}))^2 +\right.\\
&\left. (c^{(1)}_{\text{out}} - c^{(2)}_{\text{out}})^2 \tan^2\left( \pi \, \frac{m}{q} \right) \right).
\end{split}
\end{equation}
This form makes it easy to get the two largest eigenvalues, which leads to the final form of Eq. (\ref{eq:max_lambda}).

\end{document}